\def\gtorder{\mathrel{\raise.3ex\hbox{$>$}\mkern-14mu
    \lower0.6ex\hbox{$\sim$}}}
\def\ltorder{\mathrel{\raise.3ex\hbox{$<$}\mkern-14mu
    \lower0.6ex\hbox{$\sim$}}}

\documentclass[
    ,final            
  ]
  {aipproc}

\layoutstyle{6x9}

\begin{document}

\title{Modeling Dynamics in the Central Regions of Disk Galaxies}

\author{Isaac Shlosman}{address={
  Department of Physics \& Astronomy, University of Kentucky,
  Lexington, KY 40506-0055, USA}
}

\begin{abstract}
The central regions of disk galaxies are hosts to supermassive black holes
whose masses show a tight correlation with the properties of surrounding
stellar bulges. While the exact origin of this dependency is not clear,
it can be related to the very basic properties of dark matter halos
and the associated gas and stellar dynamics in the central kpc of host
galaxies.
In this review we discuss some of the recent developments in modeling
the wide spectrum of dynamical processes which can be affiliated with the
above phenomena, such as the structure of molecular tori in AGN, structure
formation in triaxial halos, and dissipative and non-dissipative dynamics 
in nested bar systems, with a particular emphasis on decoupling of gaseous
nuclear bars. We also briefly touch on the subject of fueling the nuclear
starbursts and AGN.
\end{abstract}

\maketitle


\section{1. Introduction}

There is a natural tendency to extrapolate known conditions and evolution 
beyond their original limits. A somewhat similar situation has happened in 
galactic dynamics when processes operating on large spatial scales have been 
postulated to play a similar role within the central kpc with only a minor
correction for shorter timescales. However, during the last decade it became
apparent that a number of fundamental differences exist between galactic
disks at large and their central regions. These differences provide an ample
evidence that the traditional roles of various dynamic agents can change in the
very galactic interiors. The corollary is that a large dynamic range, even
when limited to within the central kpc only, results in a plethora of
phenomena so diverse in their nature that various new techniques are required
to address the overall evolution in the region. In other words, the central
kpc of disk galaxies is an
entangled web of complex processes whose understanding will depend on high
resolution observations and state-of-the-art numerical modeling, and no single
observational or modeling technique can provide an adequate description
of the
processes thought. Here we review some of processes thought to dominate 
the region from a fraction of a pc to a kpc --- about 4 orders of magnitude in 
spatial scale, and describe the present status of their modeling. What are
the most important factors within this region that can influence its
evolution? 

First, the central kpc of a fully-grown galactic disk appears to be dominated
by the baryonic matter 
(e.g., McGaugh, Rubin \& de Block 2001;
de Block \& Bosma 2002; Bosma 2002), while beyond the $\sim 10$~kpc scale the
dark matter (DM) prevails. Universally obtained {\it numerical} cuspy
density profiles in CDM halos (e.g., Navarro, Frenk \& White 1996,
hereafter NFW) could
be flattened during a stage of inhomogeneous baryonic inflow toward the
central regions (El-Zant, Shlosman \& Hoffman 2001). It is possible that the
size of the flat core in the DM halo has evolved in conjunction with the
growing galactic disk and is heavily influenced by the merger history.  How
much dissipation has been involved in the formation of inner disks and bulges
has yet to be determined, but it seems highly unlikely that their origin was
purely collisionless. 

Second, the central supermassive black holes (SBHs) situated in galactic 
centers serve as another defining characteristic of these regions. The SBHs 
appear ubiquitous and the observed tight correlation between their masses and
the dispersion velocities in the galactic bulges has verified unambiguously
that
host galaxies are aware of their central constituents on the smallest spatial
scales, and vice versa (e.g., review by Ferrarese \& Ford 2004). This
correlation is strikingly at odds with the small
radius of influence of the SBH, $r_{\rm SBH}\sim 5 M_{{\bullet},7}
\sigma_{100}^{-2}$~pc, where $M_{{\bullet},7}$ is the SBH mass in units of
$10^7~{\rm M_\odot}$ and $\sigma_{100}$ is the velocity dispersion in the
bulge in 100~${\rm km~s^{-1}}$. Note, that the SBH interaction with the
central kpc of galaxies is twofold: radiative (when it is in the active
galactic nucleus [AGN] phase) and gravitational (section~3).

Third, large-scale galactic disks often harbor rapidly rotating stellar 
bars --- those which extend to their corotation radii (e.g., Athanassoula
1992). The corresponding properties of nuclear bars are much less known, but
it is clear that they can be either star or gas dominated (e.g., review by
Shlosman 2001), and are not necessarily fast rotators (section~4). Because bars
frequently have a decisive effect on disk dynamics and evolution, we shall pay
special attention to bar-induced dynamics in this review.

Next, the surface density and the amounts of cold (molecular) gas generally
increase towards the galactic center, and this trend is more profound in
barred galaxies (Sakamoto et al. 1999; Jogee, Scoville \& Kenney 2004).
Moreover, there are clear indications that the state
of molecular gas in the center differs from that in the outer disk in such
a way that the fraction of diffuse (unbound) molecular gas is much higher,
and its dense part is both denser and warmer (e.g., H\"uttemeister 2002). 
The gas at the very center of AGN host galaxies, the inner few pc from the 
SBHs, is detected in the form of molecular tori (e.g., Antonucci 2002) and a
number of alternative explanations to the existence of these geometrically
thick configurations are known. We discuss these issues in section~2, and only 
note here that the vertical structure of these tori can be heavily 
influenced by magnetic fields amplified in the underlying sheared accretion 
flow and modulated by the AGN radiation field. 


The structure of this paper is as follows. In the next section we discuss the
gas dynamics on the smallest relevant scales, $\sim$~pc. Section~3 analyzes the
SBH-bulge relation within the context of triaxial background potentials.
New developments in nested bar dynamics are addressed in section~4, and
some aspects of AGN fueling are given in section~5. Finally, section~6 deals
with the possible causal connection between the nuclear starbursts and AGN.

\section{2. Modeling Molecular Tori in AGN: Disk Starbursts $vs$ Disk Winds}

The central engine in AGN is surrounded by a dusty torus and is visible
only in the pole-on viewing of ``type 1'' sources and blocked from view in
edge-on ``type 2'' sources. Compelling evidence for the toroidal
orientation-dependent obscuration comes from both the optical, especially line
spectropolarimetry (e.g., Antonucci 2002), and X-ray regimes. The torus reveals 
itself also in dust IR emission, including the recent direct $K$-band imaging
of the inner region in NGC~1068 (Weigelt et al. 2004) and is likely to consist
of
a large number of individually very optically thick dusty clouds (Krolik \&
Begelman 1988). A serious impediment to detailed dynamical calculations of the 
torus structure has been the lack of guidance from realistic radiative transfer 
models based on IR observations. A formalism to handle dust clumpiness has been
developed only recently, and shows that it resolves fundamental
difficulties encountered by previous theoretical efforts (Nenkova, Ivezic
\& Elitzur 2002; Elitzur, Nenkova \& Ivezic 2004). Clumpy models have since
been employed in a number of observational studies, including that
of {\it Spitzer} observations by the GOODS Legacy project (Treister
et al. 2004).

Beyond the original weakly self-gravitating cloud model of Krolik \& Begelman
(1988), two alternative scenarios have been promoted in order the explain the
geometrically-thick molecular tori in AGN --- the hydrostatic model where the
vertical `puffing' is performed by supernova (SN) injected energies (Wada \&
Norman 2002, Wada \& Tomisaka 2005), and the accretion disk wind model with a
combined MHD/radiative driving force (Emmering, Blandford \& Shlosman 1992;
K\"onigl \& Kartje 1994; Bottorff et al. 1997; 2000; Kartje, K\"onigl \&
Elitzur 1999). 

Numerical simulations which have 
been modeling such SN explosions as an energy source (Wada \& Norman 
2002) require an oversized torus of $\sim 100$~pc, which exceeds the upper
limits on the mid-IR sizes of $\sim 30$~pc in both NGC~1068 (Bock et al. 2000)
and NGC~4151 (Radomski et al. 2003). In such a model, much of the IR emission
will come from outer boundary, while observations favor the inner boundary
dominated tori. The starbursts in the composite Seyfert~2 nuclei discussed in
section~6 are even more extended and clearly are not associated with the tori.
In the hydrostatic model, the bulk of the energy in the torus is found in the
cloud motions and the problem lies in the efficient cooling by cloud
collisions rather than in insufficient viscous heating.

\begin{figure}[ht]
\includegraphics[clip,width=0.44\hsize]{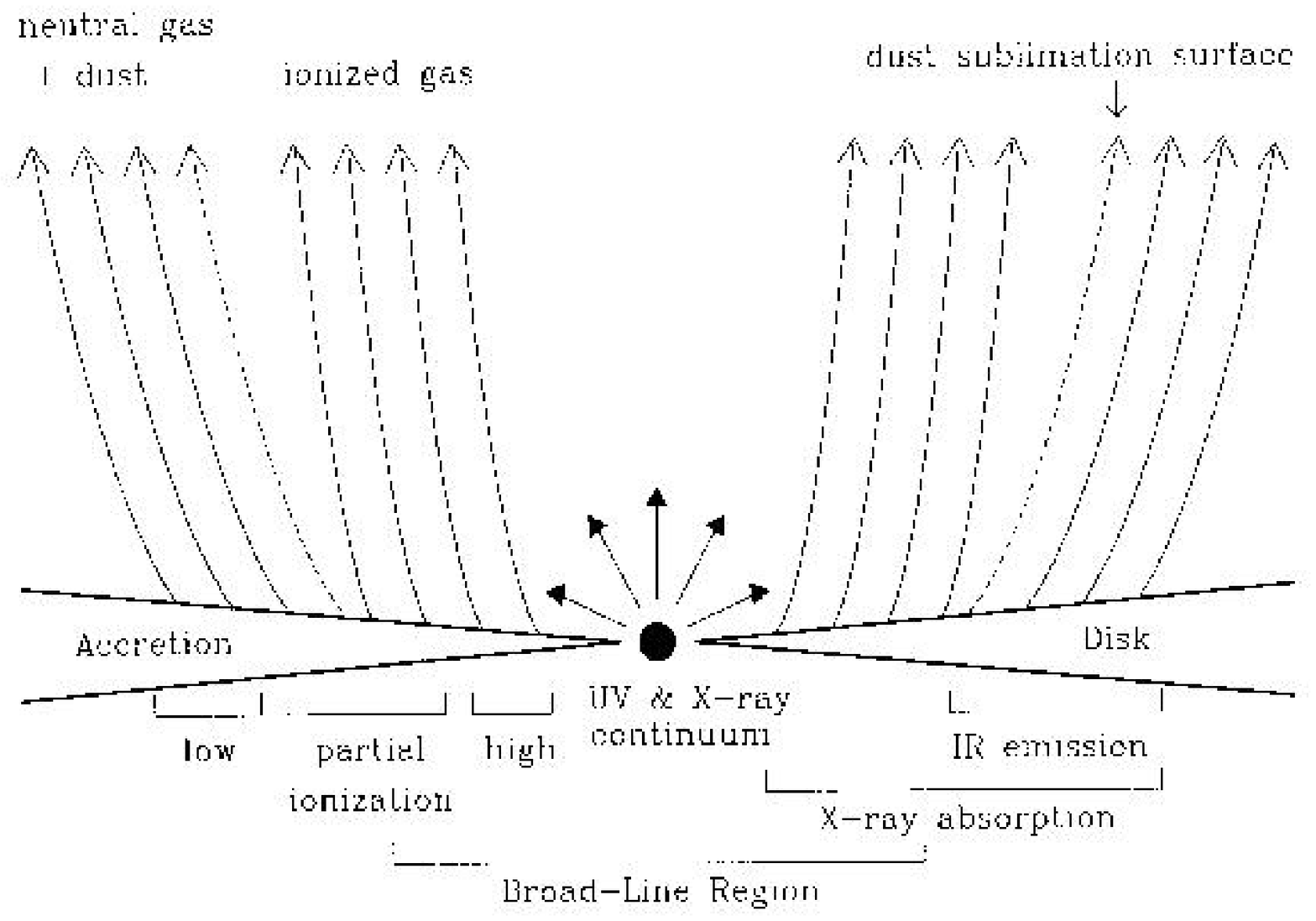}
\hfill
\includegraphics[clip,width=0.40\hsize]{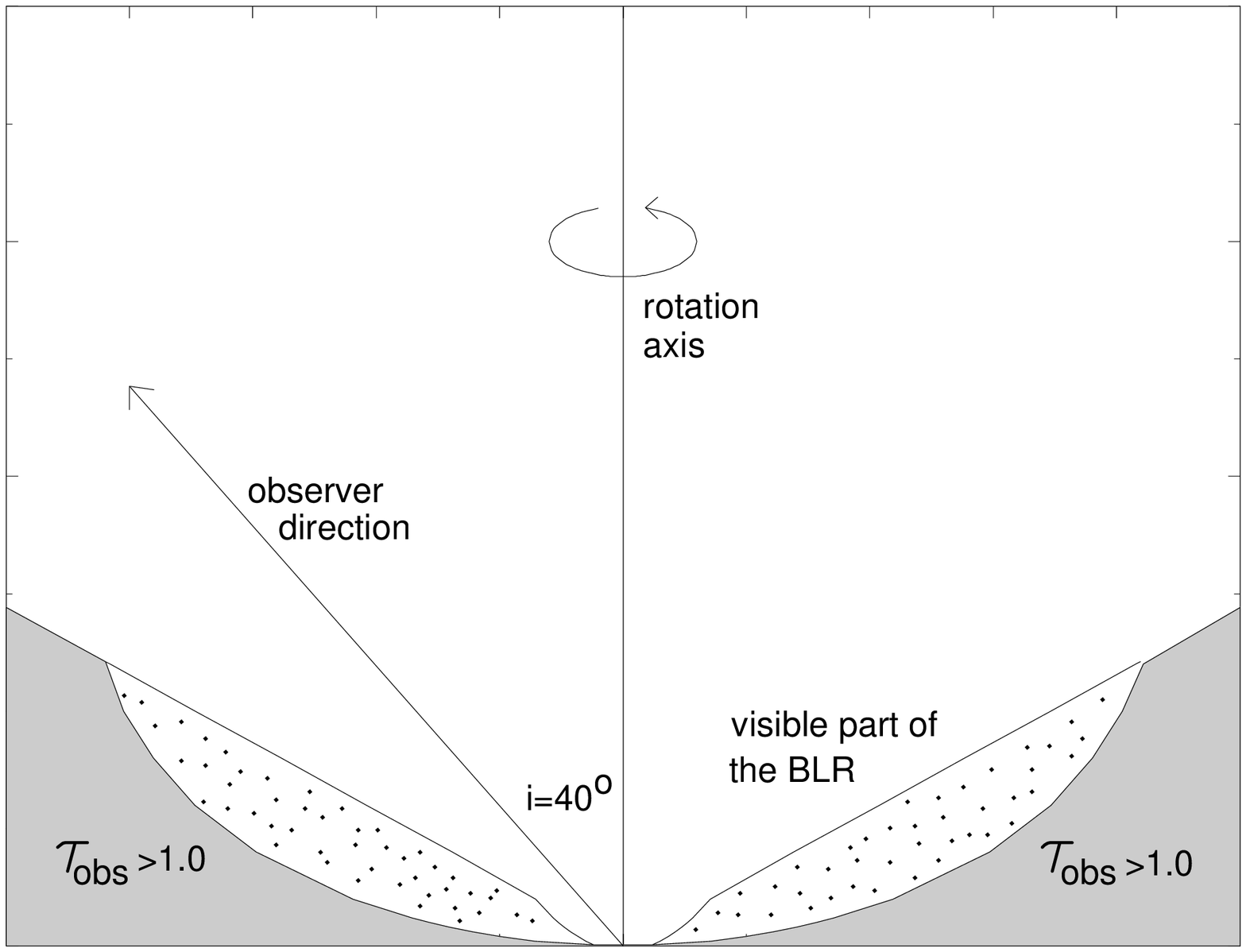}
\caption{Molecular tori from disk hydromagnetic wind in AGN (obscuring torus is
the gray-shaded region where $\tau_{\rm obs} > 1$). {\it Left:} from Kartje \&
K\"onigl (1994). {\it Right:} specific application to NGC~5548 (from Bottorff
et al. 1997).
} 
\end{figure}

On the other hand, winds fit naturally within the theoretical
framework for AGN (Fig.~1). Bipolar structures are associated with
accretion disks on {\it all} scales, from young stellar objects to quasars, and
it is natural to assume that they are symbiotically related to the dynamical
outflows from these objects. Most important, it is entirely possibly that {\it
these outflows provide a natural channel for the angular momentum loss by the
accreting material} and hence can be naturally affiliated with magnetic fields
(Blandford \& Payne 1982; Emmering et al. 1992). Such outflows
from accretion disks in AGN can possess substantial optical depths, depending
on their ability to uplift the cold material, and are clumpy. The optically
thick part of the wind meets all the requirements of the AGN torus, suggesting
itself as a natural alternative to the problematic hydrostatic paradigm. Note
that strong radiation fields generated in the vicinity of the SBH are
inevitably dynamically important, and accretion disks are prime sources of
substantial magnetic fields as well (Arav, Shlosman \& Weymann 1997 and refs.
therein). It seems increasingly plausible that magnetic fields which provide
the source of viscosity within the AGN disks (e.g., Hawley, Gammie \& Balbus
1996; Hawley, Balbus \& Stone 2001), can also serve as a channel for angular
momentum loss by these disks to the ambient medium (Blandford \& Payne 1982;
Emmering et al. 1992; Contopoulos 1995). Moreover, the amount of radiation
momentum  ``missing" in the absorption troughs
of BAL QSOs is of the same order as the estimated flow  momentum, suggesting
that radiation pressure probably also plays a key role in accelerating  the
outflow. Lastly, at least in the BAL QSOs the absorbing column in X-rays and UV
consists  of intrinsically clumpy material (e.g., Mathur, Elvis \& Singh 1995;
Shields \& Hamann 1997) and its column density is comparable with that of the
torus (Mathur et al. 2000). Elitzur \& Shlosman (in preparation) have combined 
the wind kinematics with the IR radiative transfer, thus constraining the
parameter space of possible solutions and providing a critical comparison with
the latest observations --- the characteristic size of the torus, $\sim
10$~pc, appears to be consistent with mid-IR observations of tori in nearby
Sy~2 galaxies.
 
\section{3. Modeling $M_\bullet-\sigma$: Dynamics in Triaxial Potentials}

This correlation provides the strongest evidence so far that the SBHs in 
galactic nuclei are fueled via global sources rather than the local ones.
However, the details of this process are obscure: one can roughly 
divide the possible explanations of this `conspiracy' into gasdynamical
or stellar dynamical, and it is entirely possible that the fully
self-consistent model should incorporate both frameworks. However,
here we put the emphasis on the stellar dynamics because it allows
to relate the properties of the DM halos with those at the
smallest spatial scales in the vicinity of the SBHs where most
probably the gas dynamics in combination with radiation pressure plays 
a more dominant role.  

Halos identified in cosmological simulations are invariably found to be
triaxial (e.g., Warren et al. 1992; Cole \& Lacey 1996) with density median
axial ratios of $0.5-0.6$. Their asymmetries are expected to be partially
`washed-out,' since it has been shown (e.g., Dubinski 1994) that the settling
of a baryonic component can significantly reduce the initial nonaxisymmetry
born of dissipationless
collapse. From an observational standpoint, radial profiles of
the DM halos in fully formed galaxies tend to have nearly constant density
cores (e.g., Flores \& Primack 1994; Burkert 1995; Kravtsov
et al. 1998; Boriello \& Salucci 2001; de Block \& Bosma 2002). A similar
effect has recently been inferred for clusters of galaxies (Sand et al. 2002).
Theoretically, dissipationless CDM simulations of galactic halos
agree with a universal density profile $\propto r^{-\beta}$, where
$\beta = 1 - 1.5$ (e.g., Dubinski \& Carlberg 1991; Warren et al. 1992; Cole 
\& Lacey 1996; Fukushige \& Makino 1997; Moore et al. 1999; Klypin et al.
2000). 
NFW have found a fitting formula for the density profile of
DM halos, for a wide range of cosmological  models, in which the inner profile
diverges as $r^{-1}$, while the outer profile drops as $r^{-3}$. A cuspy
density profile arises inherently
from the cold gravitational collapse in an expanding universe (Lokas \&
Hoffman 2000). The CDM model, therefore, predicts that the inner density
profile of galactic scale DM halos is characterized by a density cusp while
observations of the dynamics of the central regions of galaxies imply a
core-halo structure of the DM. This controversy between observations of DM
cores and density cusps in numerical models is not a fundamental one and can
be resolved within the general context of CDM cosmology (El-Zant et al. 2001).

It is interesting that the orbital structure of the inner
regions of slowly rotating, asymmetric potentials with flat (i.e.,
harmonic) cores is dominated by box orbits (e.g., Lake \& Norman 1983;
Binney \& Tremaine 1987), which have no particular sense of circulation.  They
are self-intersecting and cannot be populated with gas.
Dissipation causes material to sink quickly toward the center. In the process, 
interaction with the triaxial harmonic core
causes the baryonic material to lose most of its angular momentum. One can
conjecture, that as the collapsing gas becomes self-gravitating, i.e., when its
density exceeds that of a background DM, runaway
star formation will follow, leading to the possible formation of a (classical)
bulge. El-Zant et al. (2003) have noticed that this also happens to be the 
criterion for the destruction of the
harmonic core and the emergence of loop (or tube) orbits, which do have a
definite sense of circulation. The role of the SBH is to contribute to the
emergence of these orbits in the very central region. It is the collusion
between the SBH and the more extended hot baryonic component in creating the
loop orbits that leads to the observed correlations, $M_\bullet-M_{\rm b}$
and $M_\bullet-\sigma_{\rm b}$.

\begin{figure*}[ht!!!!!!]
\vbox to2.0in{\rule{0pt}{3.8in}}
\includegraphics{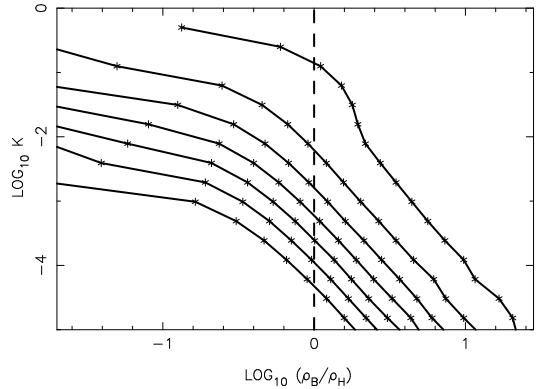}
\caption{SBH-to-bulge mass ratio $K$ for different $p_{\rm crit}$ (see text
for definition) as a function of the bulge-to-halo core density ratio required
to create closed loop orbits with axis ratios $p \ge p_{\rm crit}$ {\it
within} the bulge core, $R < 
R_{\rm b}$. The mass of the bulge core, $M_{\rm b}$ is chosen to satisfy $p >
p_{\rm crit}$ everywhere  outside the bulge.  From left to right, the
characteristic $p_{\rm crit}$ associated with a given curve increases from 0.3
to 0.9.  The bulge core density $\rho_{\rm b}$ is varied by fixing $M_{\rm b}$
and changing $R_{\rm b}$. The vertical dashed line at $\rho_{\rm b}/\rho_{\rm
H} =
1$ is the approximate boundary between the self- and non-self-gravitating
regimes in the bulge. For densities $\rho_{\rm b}/\rho_{\rm H}\sim 1$, the
$p_{\rm
crit}=0.9$ curve corresponds to an (unrealistically) overmassive SBH and is
shown for comparison only. The range of $K$ values, for these densities, thus
is limited to within $10^{-4}-10^{-2}$ (El-Zant et al. 2003).
}
\end{figure*}

Fig.~2 summarizes the SBH-to-bulge mass ratios, $K$, predicted by the model
as a function of a bulge-to-halo core density ratio, $\rho_{\rm b}/\rho_{\rm
H}$
for the halo potential axial ratio of 0.9. The seven curves shown differ in
the value of $p_{\rm crit}$ --- the critical value of axial ratios of loop
orbits
above which the gas circulation can be sustained for secular timescales. The
value of $p_{\rm crit}$ appears to be in the range of $\sim 0.4-0.5$, based on
numerical simulations of Smoothed Particle Hydrodynamics (SPH)  in a fixed
background potential, for both non- and self-gravitating gas,
with the isothermal equation of state (Pichardo \& Shlosman 2005). This
means that the model predicts a SBH-to-bulge mass ratio within the observed
range of $K\sim 10^{-4}-10^{-2}$, based on the mix of gas- and stellar
dynamical processes within triaxial DM halos.

While this approach is promising, the real challenge is to incorporate it
within the general framework of galaxy formation and evolution. Processes
like galaxy mergers and dissipative dynamics affect both the DM halo shapes
and their radial density profiles, as well as the growth of the SBHs. It is 
the tightness of the $M_\bullet-\sigma_{\rm b}$ correlation across a vast
dynamic range which is puzzling.

\section{4. Modeling Nested Bars: Dissipative and Non-Dissipative Dynamics}
 
The large fraction of stellar bars in the local universe has been noticed
already by Hubble (1936; e.g., tuning fork), and their importance for 
understanding dynamics in disk galaxies has been widely recognized (e.g.,
Athanassoula 
1984). But the exact role of bars in cosmological evolution is only now
beginning
to be unraveled. The claim that bars disappear at redshifts above 0.5 (Abraham
et al. 1999) has not been substantiated. The GEMS survey (Rix et al. 2004) has
just shown that the fraction of strong bars remains unchanged up to
redshifts of $\sim 1$ 
at least, with their size and ellipticity distributions being compatible with
a mild evolution {\it not} driven by major mergers (Jogee et al. 2004b). The
main point of interest here is that these stellar bars have an effect on the 
mass redistribution in a galaxy, specifically, they induce gas inflows to the 
central kpc, which is observed (e.g., Jogee 2004) and modeled numerically, 
both neglecting the gas self-gravity (e.g., Combes \& Gerin 1985) and
accounting for it (e.g., Simkin, Sue \& Schwarz 1980; Schwarz 1984;
Athanassoula 1992; Shlosman \& Noguchi 1993; Friedli \& Martinet 1993; Heller
\& Shlosman 1994). Typically, the most direct dynamical effect bars have on the
host galaxy is by means of gravitational torques and resonances.
Unfortunately, the bar torques 
have little effect within the central kpc where the underlying gravitational 
potential becomes nearly axisymmetric. The outer inner Lindblad resonance
(OILR), if this exists, usually borders this region and the observed nuclear
spirals (e.g., Laine et al. 1999) which can be driven by the bar are too weak
to trigger any inflow (Englmaier \& Shlosman 2000). 

This is why the existence of a different species --- nuclear bars, confined
within the inner kpc region, is so important for the evolution of galactic
centers. In this section we review the few possibilities which lead to 
a nonaxisymmetric mass distribution in these regions and focus on the 
gas inflow.

\subsection{4.1. Stellar and gaseous secondary bars: observations}

Nuclear bars almost always come in conjunction with large-scale bars (Laine
et al. 2002) --- a clear
indication that this is a prerequisite for their existence, although one can
envisage a scenario where they form separately. We, therefore, follow the 
established prescription and call the large-scale bars `primary,'
and the nuclear bars --- secondary. The secondary bars can
be pure stellar, gaseous or of mixed content. Evolutionary patterns will
differ in all three cases.

Systems of two nested bars which involve a stellar secondary bar appear to be 
unexpectedly abound in the local universe --- about 1/3 of all barred disks are 
hosts to these configurations (Laine et al. 2002; Erwin \& Sparke 2002). 
Comparable statistics of gaseous or gas-dominated secondary 
bars is unavailable currently. The detection of stellar nuclear bars is limited
to 
optical and NIR photometry so far. Laine et al. (2002) have argued that this is
a clearly insufficient method because it is based
on the fitting and isophotal analysis which can be affected by, e.g.,
nuclear clusters of star formation, or
dust extinction. Regan \& Mulchaey (1999)  and Martini \& Pogge (1999) have 
suggested to use the offset dust lanes --- which are characteristic of
large-scale bars, to detect nuclear bars. However, the dynamics of stellar
secondary bars in these systems differs substantially from their large-scale
counterparts (section~4.2) and no large-scale shocks will form under these
conditions (Shlosman \& Heller 2002).

Laine et al. (2002) find that the distribution of normalized (to $D_{25}$) 
stellar nested bar sizes is bimodal. The minimal overlap
between the distribution occurs at the (normalized to the galaxy size) bar
length $\approx 0.06$.
However, while primary bar sizes exhibit a roughly linear correlation with
the parent galaxy sizes, the secondary bar sizes are independent from the
sizes of their host galaxies.  The importance of this result can be inferred
from the fact that only in this case the normalized bar lengths will preserve
the identity of both bar groups and there will be no further mixing between
the primary and secondary bars in the normalized size space. This bimodal
distribution can be understood within the framework of disk resonances,
specifically the ILRs, which are located where the gravitational potential of
the innermost galaxy switches effectively from 3D to 2D. This conclusion is
further strengthened by the observed distribution of the sizes of 
nuclear rings which are dynamically associated with the ILRs. This could be the
first observational evidence that the ILRs play the role of dynamical
separators in disk galaxies. 

While the gas contents of nuclear bars vary, in some cases, the cold gas can be 
dynamically important, as evident from the 2.6~mm CO emission
and NIR lines of H$_2$ (e.g., Ishizuki et al. 1990; Devereux, Kenney \& Young
1992; Forbes et al. 1994; Mirabel et al. 1999; Kotilainen et al. 2000;
Maiolino et al. 2000). Such nuclear {\it gaseous} bars have no large-scale
counterparts in the local universe.

\subsection{4.2. Dynamics and dynamical decoupling of nested bars}

What makes the nested bars a truly unique dynamical system is their ability to
be in coupled or decoupled states --- the astrophysical counterpart of a
coupled oscillator which is frequently invoked in studies of nonlinear behavior
(e.g., Lichtenberg \& Lieberman 1995). The former state is characterized by 
identical pattern speeds of the primary and secondary bars, latter 
one --- by substantially different pattern speeds. We note, that the
decoupled states in galactic dynamics are not confined to nested
bars only, but rather manifest the general importance of nonlinear physics
in e.g., studies of spiral structure, interactions between the halos
and the disks, and more. Although predicted
theoretically and subsequently verified by numerical simulations, observations
provide solid support for this idea. Friedli et al. (1996) have noted that
the angle between the stellar nested bars is random, meaning that the observed
bar systems are indeed found in the decoupled state. Furthermore, the first
direct measurement of the pattern speeds in such bars has confirmed that these
are different in NGC~2950 (Corsini et al. 2003).

The main question concerning dynamics of nested bars is what defines the 
pattern speed of the secondary bar after decoupling? The details of the
actual decoupling process remain obscure. Finally, how efficient are these
systems in channeling the gas inward and fueling the nuclear starbursts and
AGN (sections~5, 6)? We, therefore, aim to
verify the basic facts about nested bar dynamics in order to understand
their potential ability to influence the overall dynamics in the central kpc.

\noindent\underline{\it Decoupling: pure stellar bars.\/}
Resonances play an important role in linear and especially in nonlinear
dynamics. Typically, they increase the dissipation rate in the gas. But a less
trivial and purely nonlinear effect is mode coupling. This applies to
spiral and bar modes (e.g., Tagger et al. 1987). The basic idea is that
nonlinear modes can exchange energies and angular momentum and hence can
support pattern speeds which otherwise would decay exponentially. The
question that interests us here is to what degree the nonlinear mode
coupling can be responsible for the initial separation and the subsequent
maintenance of different pattern speeds in nested bar systems. 
 
Numerically, the decoupling of pure stellar bars is achieved through initial
conditions only (e.g., Friedli \& Martinet 1993). With a sufficient number of
particles, this state is long-lived. The pattern speed of the
secondary bar puts its corotation radius at the position of the ILR of
the primary bar. This pumps energy into the swing amplification
cycle of the secondary bar. It is not clear, however, how the initial condition
necessary for decoupling in this case can be achieved in the normal process of 
galactic evolution --- as we shall see, the crucial ingredient, the gas, which 
is responsible for dissipation is missing in these calculations.

The real stellar nested bar systems are so abundant in the local universe that
they must be long-lived. We discard the possibility that such systems can be
built mostly from trajectories which transit between the two component bars.
El-Zant \& Shlosman (2003) have shown that a typical nested bar system is made
of regular orbits confined to each bar and trapped 
chaotic orbits in the vicinity of the regular regions. But what are the limits
in the parameter space corresponding to such a stable system?

\begin{figure*}[ht!!!!!!]
\vbox to3.6in{\rule{0pt}{3.8in}}
\includegraphics{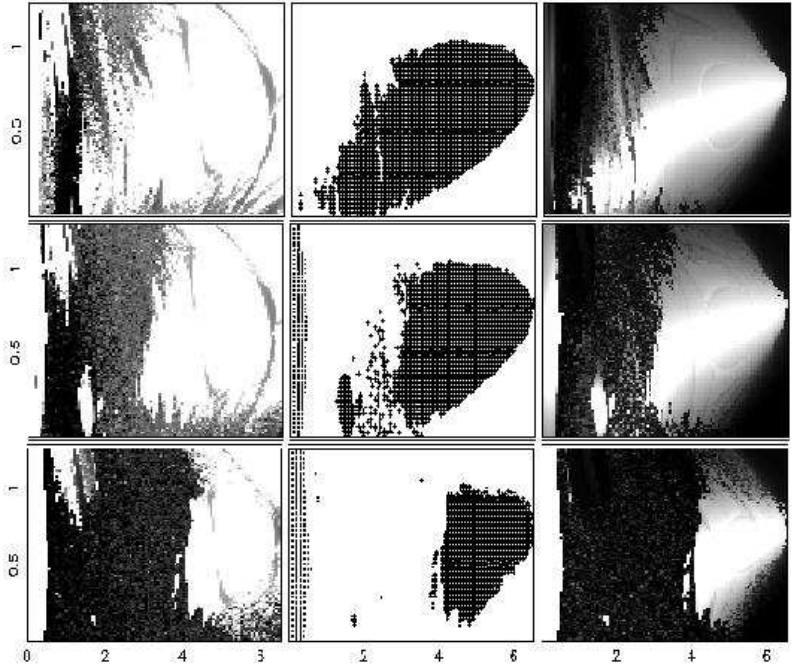}
\caption{Double-bar galaxy models with secondary bar masses increasing from top 
to bottom by a factor of 4 each. The middle panels represent the generic model. 
The abscissa refers to radii (in kpc) and the ordinate to fractions of the
local (circular) rotation velocity. \underline{\it Left column:\/} Grayshades
show the logarithms of Liapunov exponents (black-to-white corresponds to
increased stability, from characteristic Liapunov time of $10^2$~Myr, i.e.
chaotic orbits, to $10^4$~Myr, i.e. regular orbits). 
\underline{\it Middle column:\/} Exhibits orbits supporting primary (crosses) 
and secondary (dots) bars, respectively. \underline{\it Right column:\/} Maps
the axial ratios, $p\equiv b/a$, of orbits (black-to-white grayshades
correspond to increase from $p \le 1$ to $p \ge 3$) (El-Zant \& Shlosman
2003).
}
\end{figure*}

The middle horizontal panel in Fig.~3 shows a generic model of nested bars
which
appears to be stable, based on an analysis of the Liapunov exponents.
Almost all orbits corresponding to the secondary bar are aligned with its major
axis (middle vertical column) and appear regular (left column), most having 
axial ratios $p \ge 3$ (right). The top panel has a 4 times less massive
secondary bar, and bar surface density ratio of only 2, while most of the
parameter space exhibits regular and trapped trajectories within the primary
bar. The secondary bar did not produce orbits required 
to sustain it, and no orbits aligned with it have been found for this model.
Simultaneously, at the bottom panels, the chaotic region just outside the inner
bar (bar-bar interface) expands, decreasing orbital support for the primary
bar by no longer displaying significant alignment with it.
This defines the critical mass fraction of the secondary bar, a few percent
of that of the primary bar. Below this mass (and surface density) the inner bar
is not
sustainable, and above --- the outer bar support is dramatically weakened. On
the basis of the left and middle panels of Fig.~3, we can safely rule out the
top and the bottom panels, thus limiting the `window of opportunity' for
the stable existence of these systems.

The majority of the trajectories aligned with the primary bar and virtually all
those aligned with the secondary bar are parented by families analogous to the
single-periodic $x_1$ family in time-independent single-barred systems, some
by  higher order families. The symmetry of these orbits requires that one
of the coordinates is maximal when the other is null (a variation of 10\% on
the exact values was introduced to allow for the effect of the perturbing
bar). This has been verified by recording the $y$-coordinate of
maximal $x$-excursion and {\it vice versa}. Between 1/3 and 2/3 orbits (from
the
top to bottom panels in Fig.~3) of trajectories have been found to be
quasi-periodic
regular ones, with exponential timescales of the order
of a Hubble time or larger. They represent  about 50\% of trajectories in the
generic model. Most ($50\%-70\%$) of the regular trajectories are elongated in
the direction of either bars, when viewed in the relevant  frame.

However, a significant fraction of the trajectories supporting the bars
includes trapped chaotic orbits. This is especially true in the case of the
primary bar, where such trapped orbits
constitute about 16\% of the supporting trajectories in the generic model.
Their fraction peaks at 20\% when the secondary bar mass is halved, while they
are quickly replaced by ``strongly
chaotic'' orbits when the mass is doubled. Although the trapped trajectories
have a non-zero Liapunov exponent, many of them mimic bar-supporting orbits
for a Hubble time or so. In general, trapped chaotic
trajectories may wander intermittently between regular and chaotic phases with
a distribution described by non-standard statistics (Zaslavsky 2002). If the
initial conditions are such that a significant number of these trajectories
are in a trapped phase, they may be of crucial importance to building such
systems as double-barred galaxies.  

\noindent\underline{\it Decoupling: non-self-gravitating gaseous bars.\/}
Gaseous bars have attracted much less attention than their stellar
counterparts, primary and secondary. The simplest explanation is that gas
rarely dominates the dynamics of stellar disks beyond the local Jeans
instability. In addition, numerical modeling of gaseous bars requires a large
dynamic range, as shown below --- the gaseous bars quickly self-destruct
when they become self-gravitating. However, it is exactly this fast process
which is of great interest in studies of fueling the central activity
whether of starburst of AGN type.
 
We start with a simplest case by adding a low surface density (i.e., 
non-self-gravitating) gas to a galactic disk with a single large-scale
bar and a double ILR. The gas response leads to the formation
of nuclear rings between the ILRs (or inside the ILR, if only one
is present). One should keep in mind also that the torques change their sign
at the IILR. The periodic orbits between the ILRs (or inside the single ILR),
so-called $x_2$ orbits\footnote{Here we mean fully nonlinear orbits, not in the
epicyclic approximation, as they have been introduced originally} (Contopoulos
\& Papayannopoulos 1980), are oriented perpendicularly to the primary bar and
serve as attractors to gas motions in their vicinity. We note that the gas
distribution in the region is never symmetric with respect to the primary bar
axis,
with the nuclear ring oriented in the first and third quadrants --- this leads
to a constant drain of angular momentum from the gas by the primary bar and to
a progressively more elongated ring.  

Hence it is not surprizing that a large fraction of cold (molecular) gas piles
up in ILR(s) region on the $x_2$ orbits, independently of the presence or
absence of self-gravitational effects in the gas. The simplest type of
decoupling between the gas and the primary bar is not driven by the gas
self-gravity (Heller, Shlosman \& Englmaier 2001). Instead, the gaseous ring
drifts deeper into the potential well because of internal dissipation. As
the gas crosses the IILR, it loses the attractor\footnote{Stricktly speaking,
the gas does not follow the periodic orbits, $x_1$ or $x_2$, rather those serve
as attractors. Viscous torques prevent the gas from being completely 
aligned with the periodic orbits} (i.e., the $x_2$ orbits),
which leads to a retrograde precession (in the primary bar frame) because
of the mutual bar-ring orientations. One possibility is that the gaseous
ring/bar remains trapped by the primary bar potential valley --- this results
in secondary bar libration about the major axis of the primary. More
spectacular is a complete
decoupling of the secondary and its substantial slowdown --- the gaseous bar
tumbles in the retrograde direction (in the primary frame) until it is captured
again by the primary potential. This effect is caused by the gravitational
torques from the primary bar and the ability of the secondary to adjust its
shape (ellipticity) depending on the mutual orientation of the bars. This
type of behavior can lead to a mild inflow across the IILR and associated
bursts of star formation.
 
\noindent\underline{\it Decoupling: self-gravitating gaseous bars.\/}
Modeling self-gravitating gaseous bars in nested systems is more difficult.
Decoupling was not obtained for this case in numerical simulations by
Wada \& Habe (1992) and Combes (1994).
Understandably, gas exhibits the strongest response to the galactic bar
potential which triggers its inflow toward the central regions. While the gas
accounts only for a few percent of the mass within the
luminous part of a galaxy, its contribution is expected to rise by a factor of
$\sim 10$ within the central kpc. Because of this and because the random
motions within
this gas are substantially lower than within the stellar disk, the gas can be
dynamically important within the central kpc --- a property which is
amplified even further due to the gas clumpiness. One expects the gas to 
settle on the $x_2$ orbits within
ILR(s) region in a nuclear ring or disk, whether one or two ILRs exist ---
depending on how cuspy the mass distribution is within the central kpc.   

Settling the gas on $x_2$ orbits, therefore, appears to be a general outcome
of bar-driven gas inflow towards the center. Even if no ILRs exist there
initially, the disk rotation curve which is modified by the gas gravity will
form at least one ILR. The question is how the addition of gravity will
change the gas dynamics compared with the case discussed above. Fig.~4 shows
an initial evolution which proceeds along the lines of the case for
non-self-gravitating
gas. Even the initial retrograde decoupling appears similarly because the gas
mass fraction is $\ltorder 2\%$ within corotation of the primary bar in this
model. However, this is where the evolution of both types of models diverges.
A short time after decoupling the gaseous (secondary) bar collapses
dramatically
by a factor of $\sim 7$ and accelerates in the prograde direction by about the
same factor (Fig.~5). Its pattern speed stabilizes thereafter and the
system  persists in this decoupled state. 
  
\begin{figure*}[ht!!!!!!]
\vbox to3.1in{\rule{0pt}{3.6in}}
\includegraphics{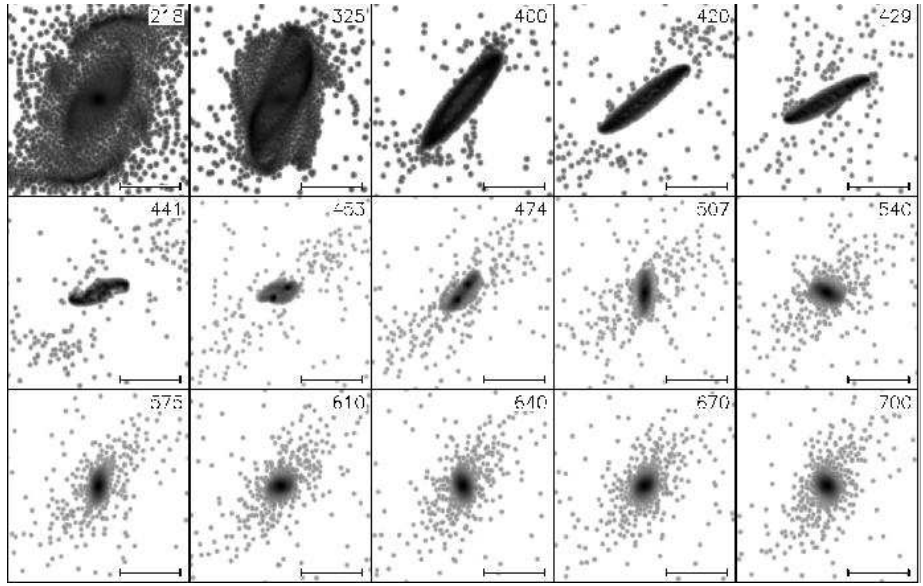}
\caption{Decoupling of self-gravitating nested bars: 6~kpc stellar (primary)
bar (not shown) and 1~kpc gaseous (secondary) bar. Each frame shows central 
3~kpc$\times$3~kpc part of a barred galaxy in the frame of the primary bar 
(horizontal). Rotation is anti-clockwise. Time in Myrs is shown in the right 
upper corners and the horizontal bar is 1~kpc in length. Gray scale represents
the gas density (Englmaier \& Shlosman 2004).
}
\end{figure*}
\begin{figure*}[ht!!!!!!!!!!!!!!!!!!!!!!!!!!!!]
\vbox to1.8in{\rule{0pt}{2.61in}}
\includegraphics{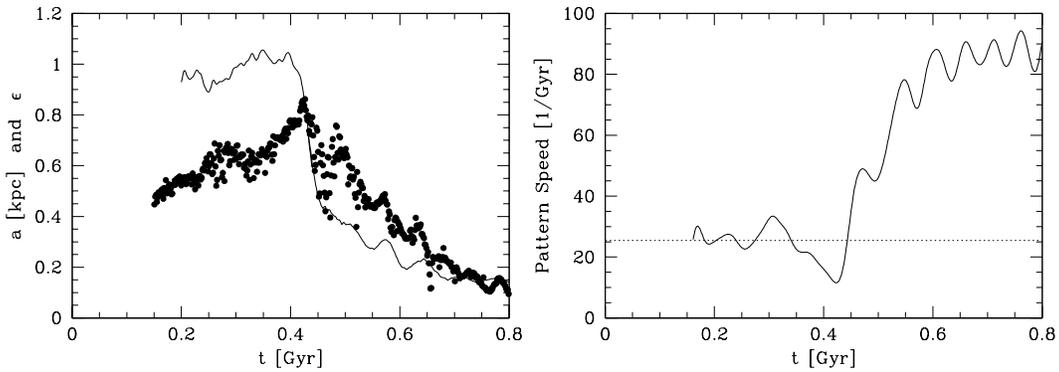}
\caption{{\it Left:} Evolution of ellipticity, $\epsilon=1-b/a$ (dotted line),
and semimajor axis, $a$ (solid line) of the secondary bar.
{\it Right:} Pattern speeds of primary ($\Omega_{\rm p}$, dots)
and secondary ($\Omega_{\rm s}$, solid line) bars in the inertial frame
(Englmaier \& Shlosman 2004).
}
\end{figure*}

Analysis of this model provides some insight into the decoupling process and
the special role of gas in it. First, the retrograde decoupling is of no
principal importance here --- an increase of $\sim 20\%$ of gas mass results in
a purely 
prograde decoupling without the retrograde phase. Second, the final steady
state in the model follows from a finite dynamic range in the numerical model
--- the limiting gravitational softening of 150~pc. This is the end-size of
the secondary bar after the catastrophic inflow and this apparently determines
its pattern speed. 

What is more important is that the secondary bar ellipticity (i.e., its
strength) increases beyond 0.8 --- its axial ratio becomes $b/a < 1:5$. This
exceptional increase in the bar strength also serves as the main reason for
its demise. In an early paper by Contopoulos (1981), the increase in the
non-axisymmetric part of the bar potential has been related to the onset
of stochasticity in the bar. More generally, stronger bars destabilize the
major families of orbits supporting the bar which become progressively chaotic.
Unstable regions associated with resonances widen and start to overlap.
The previously local chaos becomes global and the system
dissolves. The dissolution of strong collisionless bars, with axis ratio $1:5$
and below, has been modeled (e.g., Teuben \& Sanders 1985). 

The situation is expected to be even more dramatic in self-gravitating gaseous
bars because of a two-fold reason. First, gaseous bars can reach an even
smaller
axial ratio due to the dissipation. Second, the gas cannot reside on
intersecting orbits --- hence the increase in the fraction of chaotic orbits
should have catastrophic consequences for gaseous bars. Fig.~5 confirms
this overall behavior. 

The main difference between gaseous and stellar secondary bars is this ability
of the former to contract dramatically due to internal shocks. A number of
triggers which are capable to excite these shocks can be listed, however 
it is plausible that the increased bar strength and the associated growth of
the fraction of chaotic orbits is sufficient to cause this runaway. But how
does decoupling proceed? This bar strength is a direct consequence of gas
self-gravity. The test model with non-self-gravitating gas shows a diverging
evolution  --- it never decouples in the prograde direction. 

Based on a large number of models, we find that the decoupling process
approximately preserves the product of secondary bar pattern speed,
$\Omega_{\rm s}$,
and the bar size, $a$, i.e., $\Omega_{\rm s}a\sim const.$ (Englmaier, Shlosman
\& Heller 2004). This means that the decoupling gas is losing its angular
momentum, $\Omega_{\rm s}a^2$, to the primary bar. But it also means that
$\Omega_{\rm s}$
is increasing sharply during this phase. One can make an analogy between this
behavior of the gaseous bars and evolution of single stellar bars. The latter
are known to extend to their corotation (e.g., Athanassoula 1992) which
typically lies in the flat part of the galactic rotation curve, and so
$\Omega(r)\sim r^{-1}$. Pure stellar bars embedded in live halos are
transferring their angular momentum to these halos and as a result slow down
their pattern speed, $\Omega_{\rm p}$. Their growth provides a partial
`compensation' for the loss of angular momentum, hence  $\Omega_{\rm p}r_{\rm
p}\sim const.$, where $r_{\rm p}$ stands for the stellar bar size. A similar
analysis
performed in the epicyclic approximation for the secondary bars leads to the
same relation between the pattern speed and the bar size, {\it if} the bar
extends to the position of the IILR --- which is what is observed in our
numerical simulations. We note this is only possible if the gas contributes a
non-negligible fraction of the mass within its radius ($\sim 10\%-16\%$ in our
models) because only in this case the contracting gas bar pulls the IILR
inward.

Two important questions follow from the above discussion: what determines the
final $\Omega_{\rm s}$, and to what degree can the mode coupling influence the
whole process.  The speedup of the secondary {\it gaseous} bar clearly depends
on its ability to contract and, therefore, to dissipate the internal
(circulation) angular momentum. In the modeling, this is limited by the
dynamic range of the numerical scheme and to a certain degree by the equation 
of state for the gas. We have used  limiting gravitational softening of
150~pc, which determines the final semimajor axis of the bar. Of course,
further decrease in the softening value requires knowledge of the equation
of state on smaller spatial scales or an
 additional  energy source, e.g.,
associated with supernovae, to prevent the development of a 
Jeans instability.
An alternative treatment is to introduce a multi-phase ISM
(e.g., Wada \& Norman 2002). Hence, in principle, we can argue that
the contraction of  gaseous bars in nested systems can proceed to
substantially smaller spatial scales than discussed here, if the issue of
limiting softening is overcome. In principle,
unless additional physical phenomena become important, this type
of contraction will lead to exceedingly smaller and denser gas bars
with substantially higher pattern speeds and it is not clear at all if
an upper limit exists for this evolutionary stage. This behavior of
course has direct consequences for fueling of nuclear starbursts and of AGN
(section~5).

While we do not find that the mode coupling plays a role in the
runaway, it can be important in locking $\Omega_{\rm s}$ with $\Omega_{\rm p}$, 
if the bar contraction is stopped for some reason. 
On the other hand, the avalanche contraction of the gaseous bar can be related
to the self-organized criticality --- a state which is an attractor for gas
dynamics within the bar.\footnote{In this framework, an analogy can be made
between the bar strength-driven chaos and the sandpile whose slope is increased
until the sand slides off (e.g., Lichtenberg \& Lieberman 1995). } 

\section{5. Gas Flows in Nested Bars and the Fueling of AGN} 

We can distingush between two types of nested bars based on their support
of radial gas flows: those secondary bars are of mixed 
or purely gaseous types. The gas dynamics differs profoundly between these two
cases. The role of these nested systems in fueling the central activity,
thermal and nonthermal, is now being investigated by a number of research
groups, and so is their effect on the overall galaxy evolution. 

If the background potential is dominated by stars, the gas inflow
can be driven either by a time-dependent potential or by large-scale shocks
delineated by the dust lanes,
e.g., as in primary stellar bars. However, these large-scale shocks are not 
detected in the secondary bars and one can understand why. If the secondary
bars extend to their corotation --- a condition necessary for the formation of
large-scale shocks --- they will feel a strongly time-dependent potential in
the outer bar-bar interface. This region will not be able to retain the gas
injected by the primary bar and the gas would `fall through' (Heller, Shlosman
\& Englmaier 2001). If, on the other hand, the bars do not extend to their
corotation, the shocks will not form either. This of course does {\it not}
preclude the gas flows across the bar-bar interface, it makes it largely
chaotic but also correlated with the bars' mutual orientation. The gas then
settles in the inner part of the secondary bar and its future evolution
is more similar to that of gas-dominated bars (see below).

The above statement applies of course to fully decoupled bars. Before the
decoupling, when the bars corotate, the secondary bars can extend to their
corotation. But the offset shocks in the gas which settles on the $x_2$
orbits will be absent as well. 

Gas inflows in the star-dominated secondary bars can be recurrent, depending
on the external fuel supply. This means that the system can exist in the
decoupled phase for long periods of time without any gas transfer to the
center. On the other hand, in the periods of active inflow, it is 
unclear whether there will be substantial star formation, which can be
inhibited by local shear, as in primary bars. The corollary is that stellar
nested bars most probably will contain an old stellar population, which
dominates
even in the secondary bars, in agreement with observations by Friedli et al.
(1996) of a few stellar nested bars galaxies. This is an important point
and shows that in secondary stellar bars, as in large-scale primary
bars, the age of the stellar population has nothing to do with its ability
to channel gas to smaller spatial scales.

If the background potential of the secondary bar is dominated by the gas,
the inflow is triggered by the growing chaos in this bar, as we have discussed
above. The avalanche-type inflow proceeds on the crossing timescale and
the inflow rates increase with decreasing radius and can be of the order of
quasar-type fueling rates (Fig.~6). Clearly, these inflows are closely
associated with dissipation and compression in the gas. As such, they
can be accompanied by bursts of star formation.

\begin{figure*}[ht!!!!!!!!!!!!!!!!!!!!!!!!!!!!]
\vbox to2.0in{\rule{0pt}{3.0in}}
\includegraphics{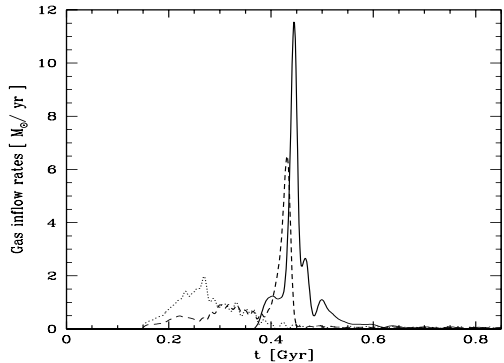}
\caption{Mass inflow rates in self-gravitating gaseous secondary bars
in decoupled nested systems. Dotted line corresponds to the inflow across
1~kpc, dashed line --- across the inner 600~pc, and solid line --- across
the central 175~pc (Englmaier \& Shlosman 2004).
}
\end{figure*}

With the current understanding that SBHs in galactic centers are
ubiquitous, at least in the nearby universe, issues related to the availability
of fuel and its delivery mechanism(s) take central stage.
The delivery mechanism is tied to the angular momentum problem (e.g., Shlosman
et al. 1989; Jogee 2004) and a number of solutions have been proposed to
resolve it. While the idea that nonlocal viscosity in the form of gravitational
torques is responsible for fueling the central acitivity, either starburst
and/or AGN-type, has acquired substantial theoretical and observational
evidence in its favor, open issues remain. In particular, within the nested
bar framework, the weak point appears to be the lack of statistics of gaseous
nuclear bars. The evolution shown here favors very short timescales of
$\sim 10^7$~yrs, which makes detection of such systems difficult. 
On the other, single bars, spontaneous or tidally-induced by mergers and
interactions, are capable of transferring the gas by only a decade in radius.
This is a clear limitation on the efficiency of large-scale stellar
bars in fueling the central activity,
and the need for additional mechanism(s) operating within the central kpc
was understood already more than a decade ago (Shlosman et al. 1989, 1990).
It is surprizing, therefore,  that Wada (2004) claims that no need exists
for triggers of nuclear activity, by e.g., counting such alternatives as
a radiative avalanche (which operates on a timescale by far exceeding the
Hubble time). We find that while the uniqueness
of the mechanism discussed here is open, solid observational and theoretical
arguments point out that the importance of nested bar systems is not only
limited to the fueling issue but is essential for the overall evolution of 
galactic disks as well.

\section{6. Composite Seyfert Nuclei: A By-Product?}

One of  the focal questions in AGN  fueling is to what  extent AGN and
nuclear starburst phenomena overlap and whether they are in any causal
relationship.  A fraction  of Seyferts, commonly  known as {\it  composite}
Seyferts, exhibit  thermal emission from a  starburst in
addition to a nonthermal nuclear component (e.g., Miley et al.  1985),
which is especially pronounced in Seyfert~2s (Edelson, Malkan \& Rieke
1987; Heckman et al.  1989). Taken in tandem with the measured excess of CO
emission, this  may hint  in the  direction of  an evolutionary
sequence:  nuclear  starburst  $\rightarrow$ Seyfert~2  $\rightarrow$
Seyfert~1 and  hence provide a ``missing" link  between starbursts and
AGN, as  first suggested by Heckman  et al. (1989), and  recently advanced by
Storchi-Bergmann et al. (2001). The key question is somewhat broader in
significance: even if there is a basis for this temporal sequence, can
we assume that the nuclear starburst is the {\it cause} of AGN
fueling and not merely a {\it by-product} of radial gas inflow? 

The recent detection of  spectroscopic signatures of hot massive stars
in a  number of Seyfert 2  nuclei has provided  additional support for
the ongoing debate on a  AGN-starburst connection (e.g., Knapen et al.
2001).  These  composite Seyfert~2 nuclei exhibit  stellar wind lines,
W-R  features and  high-order  Balmer absorption  lines superposed  on
their  ``featureless''  AGN continua.  

The  smallest of  these starbursts, when spatially resolved, have 
effective  radii between 55~pc and 200~pc (Gonz\'alez-Delgado et al. 
1998) and their morphology closely resembles that
of nuclear starbursting rings.
The phenomenon of nuclear rings in disk galaxies is fairly well
studied and understood (e.g., Athanassoula 1992; Knapen et al. 1995; Heller \&
Shlosman 1996; Knapen 2005). If indeed these
resolved nuclear starbursts are representative of all the composite 
Sy nuclei, their origin is unrelated to the central AGN. On the
other hand, it provides the clearest indication of a recent gas
inflow on the spatial scales these rings are found. So the relevant
physics to understand this phenomenon will be the galactic (gas) dynamics 
and not the AGN physics. In this case, the starburst is clearly a by-product
of gas redistribution in the galaxy.

The nuclear rings can be associated with OILR, IILR and NLR (i.e., nuclear LR)
(e.g., Shlosman 1999 for review). The OILR ring is never positioned at
the resonance but rather halfway between the OILR and IILR, while
the innermost ring sits exactly at the IILR. Numerical simulations
show that if there is more than one nuclear ring, they interact
and merge and the resulting ring appears at the IILR due to the change
of sign in gravitational torques there. This resonance is usually
found close to the center and, therefore, the observed starburst
in composite Sy nuclei can be related to the IILR. 
   
\section{7. Conclusions}

While it is clear that the issues reviewed here are not resolved completely,
substantial progress has been made on all spatial scales, from
$\sim 1-30$~pc --- the size of the molecular tori in AGN, to $\gtorder 10$~kpc
--- that of the DM halos and well beyond. A number of corollaries gradually
become obvious: the formation and evolution of galaxies appears to
be a much more correlated process compared to what had been envisioned 
before, despite the enormous dynamic range and dissipation involved.
This coupling between different parts of SBH-disk-halo systems emerges as 
strongly nonlinear, largely due to the nonlocal gravitational viscosity
involved and possibly due to mode coupling. On the other hand,
this nonlinearity results in a non-negligible contribution of chaotic
motions in building the structure of the luminous part of disk galaxies
and to a new phenomenon of dynamical decoupling of different parts
in these galaxies.


\noindent\underline{\bf Acknowledgements.\/} I am indebted to my colleagues 
for numerous and lengthy discussions
on different issues concerning this work, especially to Mitch Begelman, Ingo
Berentzen, Amr El-Zant, Peter Englmaier, Clayton Heller, Shardha Jogee, Johan 
Knapen and Barbara Pichardo. I acknowledge partial support by NASA/ATP/LTSA
grants NAG5-10823, 5-13063, by NSF AST-0206251, and by HST AR-10284. 


\bibliographystyle{aipprocl} 


\end{document}